\documentclass[10pt,journal,compsoc]{IEEEtran}

\ifCLASSOPTIONcompsoc
  \usepackage[nocompress]{cite}
\else
  \usepackage{cite}
\fi
\usepackage{cite}
\usepackage{caption}
\usepackage{amsmath,amssymb,amsfonts}
\usepackage{algorithmic}
\usepackage{graphicx}
\usepackage{color,array}
\usepackage{cite}
\usepackage{amsmath,amssymb,amsfonts}
\usepackage{algorithmic}
\usepackage{textcomp}
\usepackage[table]{xcolor}
\usepackage{url}
\usepackage{braket}
\usepackage{bm}
\usepackage{times}
\usepackage{epsfig}
\usepackage{graphicx}
\usepackage{amsmath}
\usepackage{multirow}
\usepackage{amssymb}
\usepackage{graphicx}
\usepackage{float}
\usepackage{booktabs}
\usepackage{textcomp}
\usepackage{adjustbox}\usepackage{mathtools}
\usepackage{array}
\usepackage{makecell,booktabs}
\usepackage{multirow}
\usepackage{stackengine}
\usepackage{algorithm}
\usepackage{graphicx}
\usepackage{textcomp}
\usepackage[table]{xcolor}
\usepackage{wrapfig,lipsum,booktabs}

\ifCLASSINFOpdf
\else
\fi

\begin{document}

\title{COVID-19 Pneumonia Severity Prediction using \\Hybrid Convolution-Attention Neural Architectures}

\author{Nam Nguyen,~\IEEEmembership{Student Member,~IEEE,}
        and J. Morris Chang,~\IEEEmembership{Fellow,~IEEE}
\IEEEcompsocitemizethanks{\IEEEcompsocthanksitem M. The authors are with the Department of Electrical Engineering, University of South Florida, Tampa, FL 33620.\protect\\
E-mail: namnguyen2@usf.edu}
\thanks{}}

\markboth{Hybrid Convolution-Attention Neural Architecture}%
{Nguyen \MakeLowercase{\textit{et al.}}: Titles}

\IEEEtitleabstractindextext{%
\begin{abstract}
This study proposed a novel framework for COVID-19 severity prediction, which is a combination of data-centric and model-centric approaches. First, we propose a data-centric pre-training for extremely scare data scenarios of the investigating dataset. Second, we propose two hybrid convolution-attention neural architectures that leverage the self-attention from the Transformer and the Dense Associative Memory (Modern Hopfield Network). Our proposed approach achieves significant improvement from the conventional baseline approach. The best model from our proposed approach achieves $R^2 = 0.85 \pm 0.05$ and Pearson correlation coefficient $\rho = 0.92 \pm 0.02$ in geographic extend and $R^2 = 0.72 \pm 0.09, \rho = 0.85\pm 0.06$ in opacity prediction.

\end{abstract}

}

\maketitle

\IEEEdisplaynontitleabstractindextext

\IEEEpeerreviewmaketitle

\ifCLASSOPTIONcompsoc
\IEEEraisesectionheading{\section{Introduction}\label{sec:introduction}}

The coronavirus disease 2019 (COVID-19) was declared a global pandemic by the World Health Organization in early 2020. There are $184$ million cases with approximately $4$ million deaths recorded up to July 2021 \cite{wiki}. Early detection not only ameliorates the survival rate of COVID-19 patients but also prevents the spread of diseases. Moreover, severity prediction significantly impacts the resource allocation in hospitals \cite{feng2020early,booth2021development,zhang2020viral}, which is crucial during the pandemic. Many studies \cite{fang2020sensitivity,pan2020time,bernheim2020chest,liu2020ct} shows the high correlation between severity progression of COVID-19 and the length of hospital stay, ICU admission, which is fruitful for optimal planning of follow-up medical care.

Computer-aided diagnosis based on machine learning and deep learning has become potential solutions for COVID-19 detection \cite{panwar2020application} and severity prediction \cite{cohen2020predicting,lassau2021integrating,fridadar2021covid}. The dominant solution for COVID-19 prediction is delivered through transfer learning, in which databases are abundant and adequate for the good-performed model. In contrast, severity prediction copes with extremely small cohorts, where the number of samples is inadequate to deliver well-calibrated deep learning solutions. Similar works that tackle COVID-19 severity prediction in literature are introduced \cite{cohen2020predicting,lassau2021integrating} with existing deep neural architectures.

In this work, we proposed a novel approach for COVID-19 severity prediction, which combines data-centric and model-centric improvement. We summarize our contribution as follows:
\begin{enumerate}
    \item We proposed a data-centric pre-training framework to tackle the extremely scare data scenario of COVID-19 prediction. Our proposed data-centric pre-training significantly ameliorate the performance of deep neural architectures in term of predictive power.
    \item We proposed two hybrid convolution-attention neural architectures that leverage the self-attention from state-of-the-art Transformer and Dense associative memory.
    \item The experimental results yield a noticeable improvement compared to conventional counterparts, which includes transfer learning from ImageNet and existing neural architectures. 
\end{enumerate}

The organization of our work is as follows: Section~\ref{related-works} briefly introduces related works, Section~\ref{method} gives a detailed description of our proposed approach, Section~\ref{experiment} reports our experimental design and results, Section~\ref{conclusion} gives the discussion and conclusion of our study.

\section{Related works}\label{related-works}

\subsection{Neural Architecture Design}
The design of deep neural architectures can be categorized into two approaches: (1) manual and (2) automated. In the manual design, we aim to develop the architecture of neural blocks, which requires considerable expert knowledge. For example, the residual block is introduced in \cite{he2016deep} enables more convenient optimization with residual connection from the inputs; Inception blocks enable the approximation of optimal neural spare structure by "split-transform-merge" strategy \cite{szegedy2015going}. On the other hand, automated neural architecture search (NAS) attempts to search the optimal neural architecture on a given datasets \cite{real2019regularized,zoph2018learning,nguyen2021contrastive}. The dominant approach for neural encoding for the NAS algorithm is through directed acyclic graphs, which represent blocks in neural architecture (also known as a cell). These discovered cells are then stacked to form final neural architectures.

The recent development of manual neural architecture design leverages self-attention to capture the global contextual information within the input space. The common choice for the self-attention module is Transformer \cite{vaswani2017attention}, which was originally designed for the natural language process. Vision Transformer \cite{dosovitskiy2020image} (ViT) split the input images into sequences of patches, which are taken as the input of the Transformer encoder. It is noted that in the ViT architecture, only the Transformer encoder is used. Detection Transformer leverages full architecture of Transformer for object detection tasks \cite{carion2020end}. The HybridCA \cite{nguyen2021attention} introduced learnable image representation queries for the full Transformer model, which enhances the capacity of neural architectures. CoAtNets \cite{dai2021coatnet} introduced a family of hybrid models that can improve the generalization, capacity, and efficiency.

\section{Methodology}\label{method}
\begin{figure}[t]
    \centering
    \includegraphics[width = 0.48\textwidth]{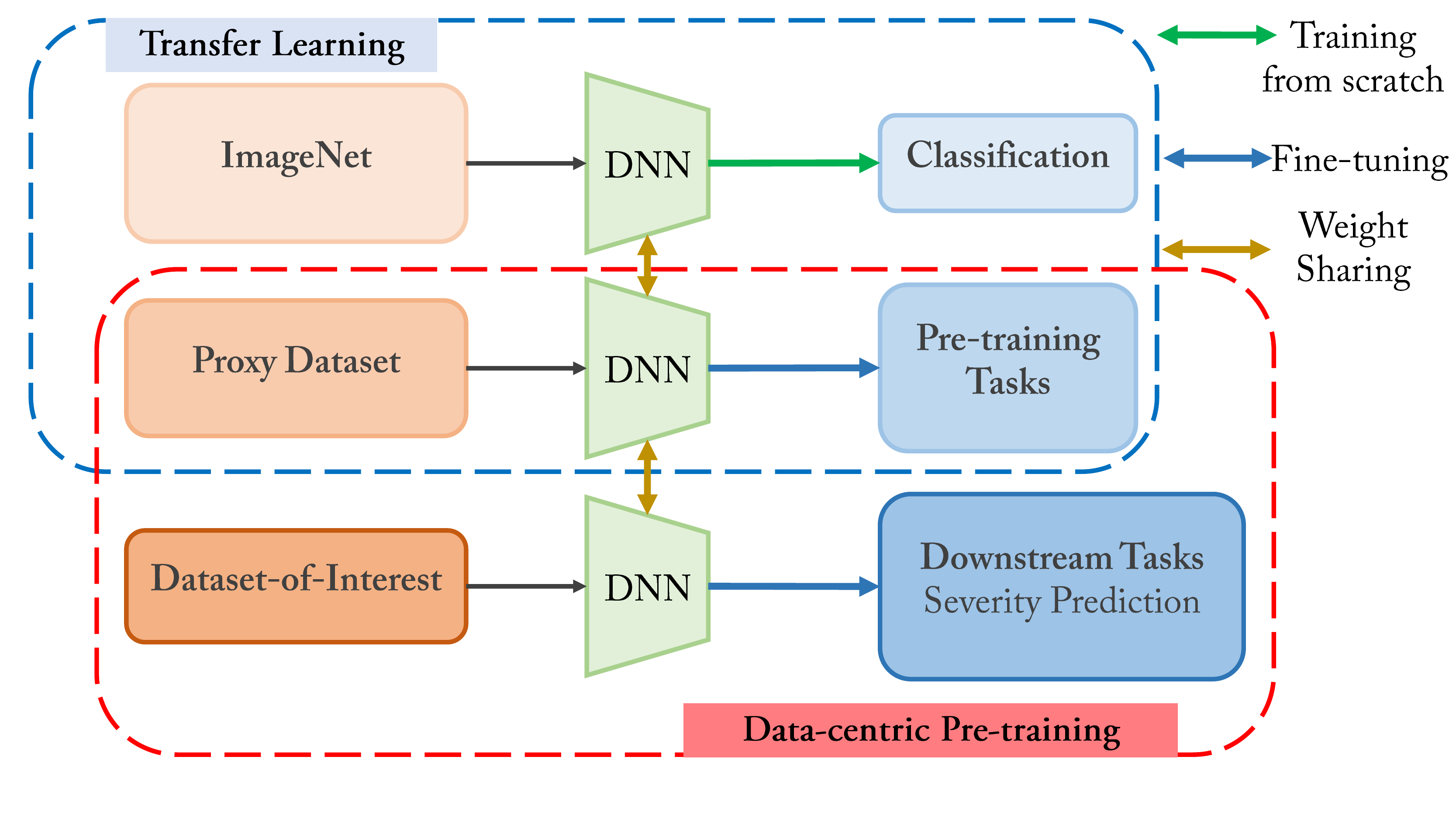}
    \caption{Illustration of our proposed data-centric pre-training framework. We highlight the difference between transfer learning and data-centric pre-training in different boxes.}
    \label{fig:data-centric}
\end{figure}
\begin{figure*}[t]
    \centering
    \includegraphics[width = 0.8\textwidth]{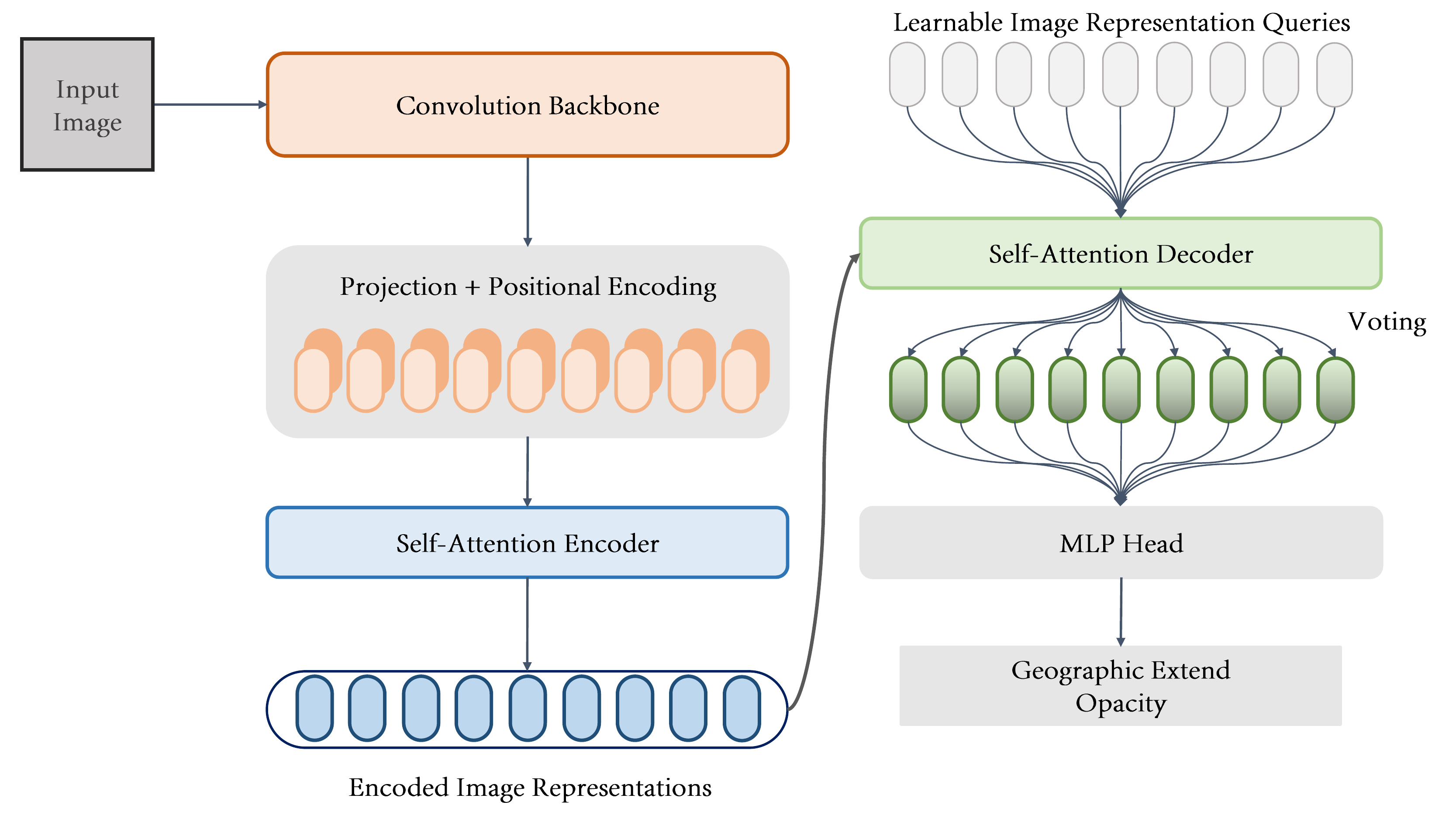}
    \caption{Illustration of Hybrid Convolution-Attention neural architecture.}
    \label{fig:hca}
\end{figure*}

\subsection{Data-centric framework for Pre-training Deep Neural Network}

The improvements of an AI system can be achieved through two main approaches: (1) model-centric and (2) data-centric development. In the model-centric approach, we aim to develop AI algorithms on given datasets, commonly fixed throughout the process. The main concentration on such an approach is delivering the optimized models for desired learning tasks, enabling state-of-the-art neural solutions. The advantage of such an approach is the convenient comparison between algorithms due to fixed collections of pre-defined datasets. However, there are several issues associated with the model-centric approach. The performance of the model-centric approach highly depends on data scenarios, which in some cases are intrinsically challenging. For example, inadequate training samples in scare data scenarios potentially lead to poor-performed models and non-robust inferences. 

In the data-centric approach, we leverage additional data for improving the performance of the AI system. The fundamental assumption of such an approach is simple but practical, which is supposed that the performance of ML/DL algorithms can be ameliorated with more relevant data samples. Transfer learning can be considered an example of the data-centric approach, which transfers knowledge from a huge domain dataset to a smaller target dataset. This approach assists in learning low-level representations from domain sets, enabling more efficient learning on target sets. However, the inherited limitation of transfer learning is negative transfer, in which domain and target datasets are irrelevant. 

Directly addressing these issues, we proposed a data-centric framework for pre-training deep neural architecture, which is illustrated in Figure~\ref{fig:data-centric}. In the pre-training phase, proxy datasets curation is required, in which proxy datasets need to be highly similar but strictly separated from the datasets of interest. The size of proxy data can be smaller or larger than the target datasets depends on desired purposed. Take automated neural architecture search (NAS) as an example, where we aim to discover the best neural solutions for a given dataset. Early works of NAS \cite{real2019regularized,zoph2018learning} search and evaluate on ImageNet \cite{deng2009imagenet} with $~14$ million samples, which leads to extensively searching time of $2250-3000$ GPU-days. Following works show that CIFAR-10 \cite{krizhevsky2009learning} ($60k$ samples) is a good proxy data for ImageNet \cite{zoph2018learning}, reducing the search time to only $1-4$ GPU days \cite{nguyen2021contrastive}. In our used case, we aim to develop a high-performance model for an extremely minimal dataset. Thus, the desired proxy dataset needs to include more samples while maintaining the similarity to the dataset-of-interest. Moreover, pre-training tasks enable learning good representations, which ensures similarity to dataset-of-interest. The pre-training tasks depend on the availability of the proxy data. In a labeled proxy, we can adopt supervised learning tasks such as classification or object detection for models to learn data representations. Regarding unlabeled proxy datasets, unsupervised and self-supervised learning \cite{misra2020self,chen2020simple} can be used to pre-train deep neural networks. 

The main objective of the data-centric pre-training phase is to help neural architecture learning good representations, which can lead to significant improvements in the downstream tasks. Good representations are expensive, which reasonable-sized representations can capture the abstraction from a vast number of inputs and mitigate the variance \cite{bengio2013representation}. The design of data-centric pre-training for COVID-19 severity prediction will be given in Section~\ref{experiment}.

\subsection{Hybrid Convolution-Attention Neural Architecture}

\subsubsection{Architecture}
We generalized the hybrid convolution-attention (HybridCA) neural architecture in \cite{nguyen2021attention}, which include two main components: (1) convolution backbone module and (2) self-attention module (Figure~\ref{fig:hca}). First, the backbone convolution transforms input features into intermediate feature maps. These image representations is then projected and vectorized inter-channel to from a collection of entities $\{ \bm{x_1,x_2,...,x_n} \}$, in which each $\bm{x_i}$ is $p-$dimensional vector in the latent space. These embedded vectors are taken as inputs of self-attention modules to extract the global contextual information and relationship amongst entities. Since the attention module possesses the permutation-invariant property, we apply the fixed positional encoding before the self-attention encoder. These encoded entities are then fed-forward into self-attention decoder and same size learnable image representation queries (IRQ), which can be considered learnable parameters of the architecture. The final prediction of HybridCA architecture is delivered by a multi-layer perceptron, customized to the desired learning tasks. 

The original architecture of HybridCA only considers the entire Transformer architecture as the self-attention module. In this work, we extend the study by investigating the effectiveness of an additional self-attention model, a dense associative memory.

\subsubsection{Transformer Model}
The Transformer is an encoder-decoder neural architecture, which contains a stack of encoder layers followed by decoder layers. In the Transformer's encoder, the core component is a multi-head self-attention block followed by a sub-sequence element-wise feed-forward network. Moreover, residual connections can be established within the encoder together with layer-wise normalization. The Transformer's decoder is similar to the encoder, except required multi-head attention for encoded representation entities. The building block for the Transformer model is multi-head attention, which is formed by the self-attention mechanism.

\textbf{Self-attention:} Given a set of image representation entities $\{ \bm{x_1,x_2,...,x_n} \}$, we denote 
\begin{equation}
    \bm{X}_{n \times p} = \begin{bmatrix}
    | & | & \dots & |\\
    \bm{x_1} & \bm{x_2} & \dots & \bm{x_n}\\
    | & | & \dots & |\\
    \end{bmatrix}.
\end{equation}
Th self-attention attempt to learn the relationship amongst input entities, producing encoded representations which captures the global contextual information from the entities. Such task requires learning three weight matrices: (1) Queries matrix $\bm{W}_{Q}^{n\times d_q}$, (2) Keys matrix $\bm{W}_{K}^{n\times d_k}$ and (3) Values matrix $\bm{W}_{V}^{n \times d_v}$. The collection of input entities $\bm{X}$ is projected onto the three learnable matrix as following
\begin{equation}
    \begin{split}
        \bm{Q} &= \bm{XW}_{Q}\\
        \bm{K} &= \bm{XW}_{K}\\
        \bm{V} &=\bm{XW}_{V}
    \end{split}
\end{equation}
The encoded representations $\bm{Z}^{n\times d_v}$ is computed as
\begin{equation}
    \bm{Z} = \text{softmax} \bigg( \frac{\bm{Q}\bm{K}^T}{\sqrt{d_q}}  \bm{V} \bigg),
\end{equation}
where $1/\sqrt{d_p}$ is temperature of the dot product in the softmax function, preventing extremely small gradients \cite{vaswani2017attention}. As a result, each element of encoded representation matrix $\bm{Z}$ is the weighted sum of all original entities in the latent space, in which weight matrix is computed by the dot-product of queries and all keys.

\textbf{Multi-head Attention} Given $B$ blocks of self-attention, multi-head attention can be formed by simultaneously computing multiple individual self-attention. We denote $\{ \bm{W}_Q^{(i)}, \bm{W}_K^{(i)}, \bm{W}_V^{(i)} \}$ for $i = 1,2,\dots B$ for each self-attention head and $\bm{Z}^{(i)}$ for each corresponding computed encoded entities. Output of multi-head attention is formed by the projection of concatenation of all elements $\bm{Z}^{(i)}$ onto $\bm{W}^{B d_v \times d}$. Hence, multi-head attention's outputs capture multiple complex interactions form projected convolution feature maps in parallel, which provide a larger receptive field.

\subsubsection{Dense associative memory}
The Dense Associated Memory (or Modern Hopfield Networks) is introduced in \cite{krotov2016dense}, which extends to continuous-valued patterns and states. The new energy function introduced in \cite{ramsauer2020hopfield} enables an exponential number of stored patterns with exponentially small retrieval errors, which is given as
\begin{equation}
    E = - \text{lse} (\beta,\bm{X}^T \bm{p}) + \frac{1}{2} \bm{p}^T \bm{p} + \beta^{-1} \log n + \frac{1}{2} M^2,
\end{equation}
where $\text{lse}(.)$ is the log-sum function, $\beta$ is the temperature, $p$ is state pattern, $n$ is the number of stored representations and $M$ is the largest norm among all stored representations. The update rule of such network by using the Concave-Convex-Procedure yields
\begin{equation}
    \bm{p}_{\text{new}} = \bm{X} \text{softmax} (\beta \bm{X}^{T} \bm{p})
\end{equation}
This proposed update rule is equivalent to the self-attention used in the Transformer model, which enables attention learning from the input data. Moreover, the new energy function with associated update rule ensure the convergence to local minimum of the energy function, leading to fast convergence.

\section{Experiments}\label{experiment}
\begin{table*}[t]
    \centering
    \begin{tabular}{c|c c c c c c}
        \toprule
        Model &  DenseNet121 & ResNet50 & EfficientNet-B1 & EfficientNet-B2 & EfficientNet-B3 & EfficientNet-B4\\
        \midrule
        Atelectasis & $0.7586$ & $0.6933$ & $0.7558$ & $0.7614$ & $0.7551$ & $0.8015$\\
        Cardiomegaly & $0.8482$ & $0.8001$ & $0.8526$ & $0.8406$ & $0.8365$ & $0.7578$\\
        Effusion &  $0.8109$ & $0.7606$ & $0.8157$ & $0.8176$ & $0.8179$  & $0.8889$\\
        Infiltration & $0.6943$ & $0.6316$ & $0.6988$ & $0.6965$ & $0.6917$ & $0.7028$ \\
        Mass & $0.7637$ & $0.7273$ & $0.7573$ & $0.7536$ & $0.7589$ & $0.7796$\\
        Nodule & $0.7081$ & $0.6554$ & $0.7149$ & $0.7106$ & $0.7196$ & $0.7324$\\
        Pneumonia & $0.7030$ & $0.6488$ & $0.6902$ & $0.6948$ & $0.6987$ & $0.7313$\\
        Pneumothorax & $0.8412$ & $0.8037$ & $0.8383$ & $0.8455$ & $0.8435$ & $0.8206$\\
        Consolidation & $0.7282$ & $0.6912$ & $0.7171$ & $0.7272$ & $0.7235$ & $0.7998$\\
        Edema & $0.8363$ & $0.8087$ & $0.8382$ & $0.8316$ & $0.8314$ & $0.8868$\\
        Emphysema & $0.8680$ & $0.8456$ & $0.8319$ & $0.8737$ & $0.8736$ & $0.8330$\\
        Fibrosis & $0.7689$ & $0.7334$ & $0.7839$ & $0.7714$ & $0.7729$ & $0.7667$\\
        Pleural Thickening & $0.7460$ & $0.7098$ & $0.7349$ & $0.7381$ & $0.7446$ & $0.7781$\\
        Hernia & $0.8097$ & $0.7331$ & $0.8225$ & $0.7972$ & $0.7829$ & $0.7444$\\
        COVID-19 & $0.9992$ & $0.9993$ & $0.9998$ & $0.9998$ & $0.9998$ & $0.9994$\\
        \midrule
        Mean AUC & $0.7923$& $0.7495$ & $0.7901$ & $0.7907$ & $0.7900$ & $0.8018$\\
        \bottomrule
    \end{tabular}
    \caption{Experimental results of pre-training phase. Details of training setting is given in Section~\ref{sec:pre-training}}
    \label{tab:pretraining}
\end{table*}

\subsection{Experimental Designs}
\begin{figure}[hbt]
    \centering
    \includegraphics[width = 0.48\textwidth]{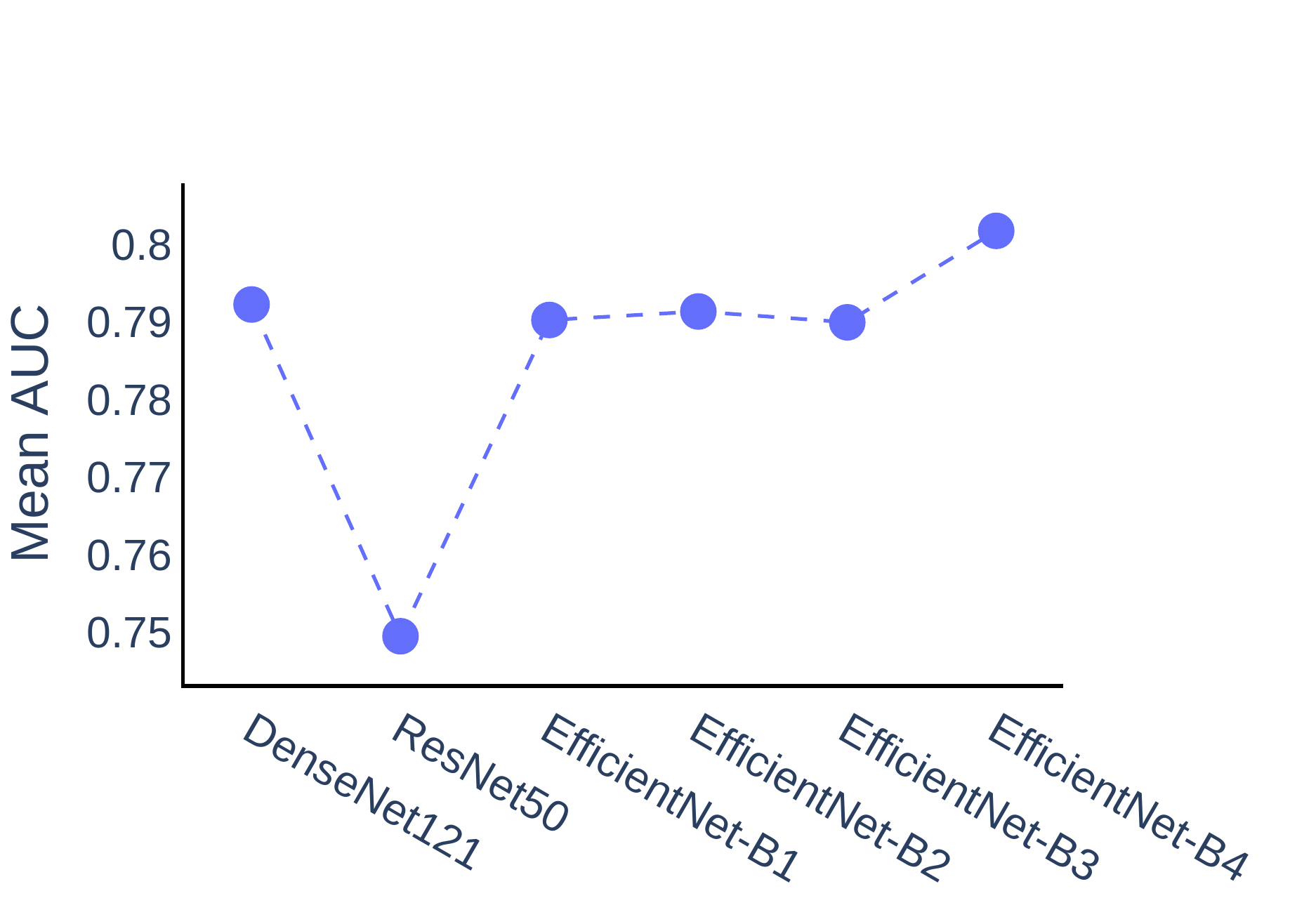}
    \caption{Experimental results of pre-training phase. The same experiment setting is used across all training sessions, which is reported in Section.}
    \label{fig:pretraining}
\end{figure}

\begin{figure}[t]
    \centering
    \includegraphics[width = 0.48\textwidth]{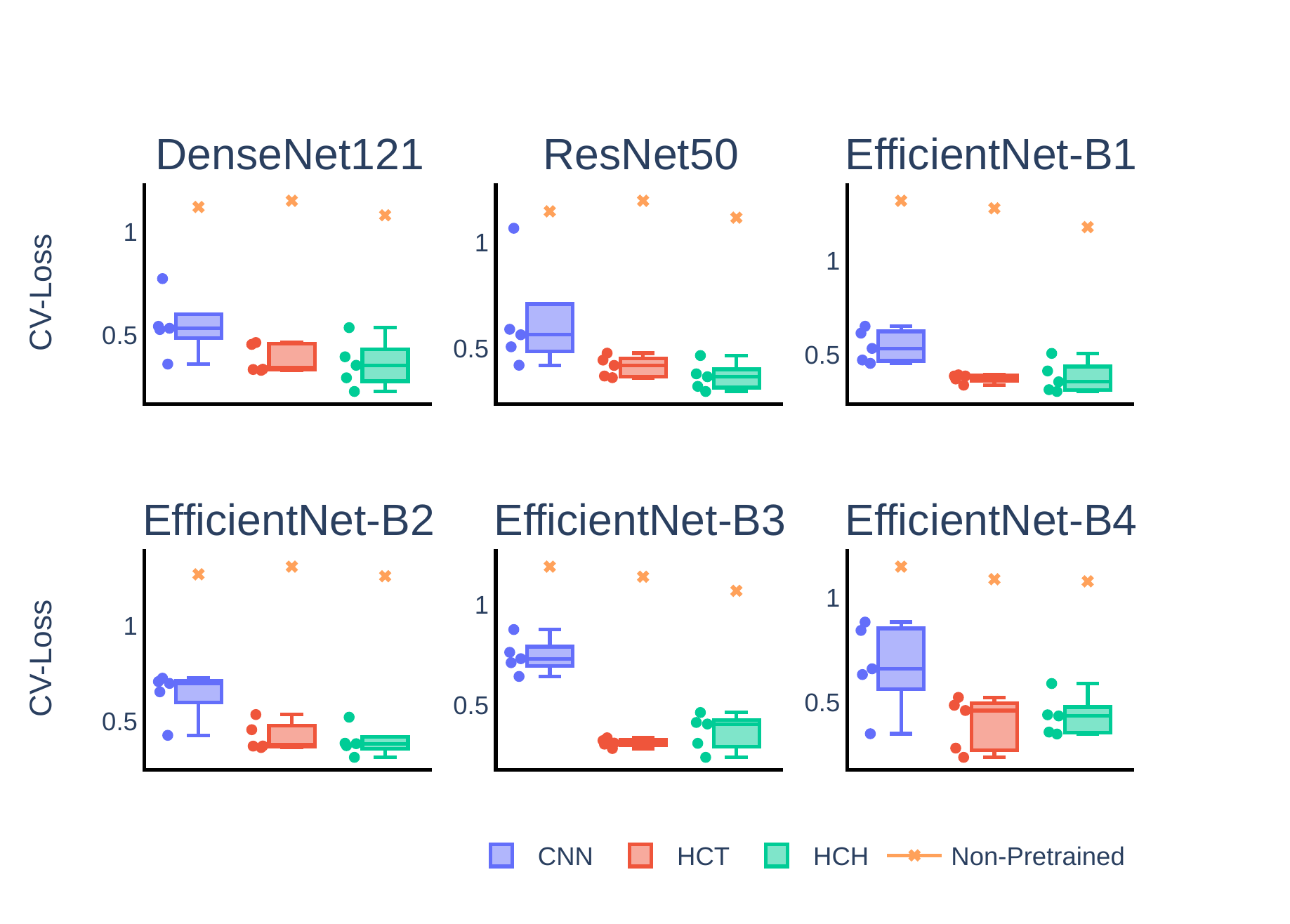}
    \caption{Effects of data-centric pre-training on training progress of different models. We report the cross-valiation loss computed by Equation~\ref{smoothL1}. The results from non-pretrain (transfer learning) is depicted in yellow.}
    \label{fig:training}
\end{figure}
\subsubsection{Design of Data-centric pre-training task}\label{sec:pre-training}

In the pre-training phase, we collect two databases for the proxy dataset of COVID-19 severity prediction: (1) NIH Chest X-ray Dataset \cite{wang2017chestx}, which include $112,120$ Non-COVID-19 X-ray images with $14$ disease labels from $30,805$ patients and (2) COVID-19 database includes $3,671$ images collected from various resources \cite{chowdhury2020can,rahman2021exploring}. It is noted that these databases are completely separated from the dataset used for severity prediction. 

The pre-training task for the proxy data is multi-label classification. The target vector (label) for each input instance is a $15$-dimensional vector ($14$ types of disease plus COVID-19 class) with binary entries, representing the presence of related diseases. In other words, the zero vector $y = [0,...,0]$  represents normal case, while an unit entry at location $C$ represents the appearance of $C^{th}$ disease. This label encoding guarantees that the model cannot infer normal and disease classes concurrently and that learning tasks are considered a regression-like problem.

Within the scope of this study, we investigate five backbone convolution neural networks with different model complexity, which are DenseNet121 \cite{huang2017densely}, ResNet50 \cite{he2016deep}, EfficientNet-B1 to B5 \cite{tan2019efficientnet}. The loss function for the pre-training task is binary cross-entropy loss. To optimizing model parameters, we use AdamW optimizer with an initial learning rate of $10^{-6}$ and weight decay $0.01$. We pre-train backbone CNN with initial ImageNet weights for $100$ epochs. We discuss the experimental results of the pre-training phase in Section~\ref{experiment}.
\begin{table*}[!t]
    \centering
    \begin{tabular}{c| c c| c c | c c | c  c}
        \toprule
        \multirow{2}{*}{\textbf{Model}} &  \multirow{2}{*}{\textbf{Pre-trained}} & \multirow{2}{*}{\textbf{Attention Module}} & \multirow{2}{*}{\textbf{MAE}} & \multirow{2}{*}{\textbf{MSE}} & \multicolumn{2}{c}{\textbf{Geographic Extend}} &  \multicolumn{2}{c}{\textbf{Opacity}}\\
        & & & & & $R^2$ & $\rho$& $R^2$ & $\rho$\\
        \midrule
        \multirow{4}{*}{DenseNet121} & -&- & $1.28 \pm 0.23$ & $2.68 \pm 0.75$ & $0.39 \pm 0.13$ & $0.62 \pm 0.1$ & $0.32 \pm 0.18$ & $0.51 \pm 0.23$\\
         & $\checkmark$ & - & $0.94 \pm 0.15$ & $1.39 \pm 0.42$ & $0.74 \pm 0.04$ & $0.86 \pm 0.02$ & $0.55 \pm 0.05$ & $0.74 \pm 0.03$\\
         & $\checkmark$ & Transformer &\cellcolor[HTML]{D3D3D3} $0.74 \pm 0.07$ & \cellcolor[HTML]{D3D3D3}$0.93 \pm 0.19$ & \cellcolor[HTML]{D3D3D3}$0.82 \pm 0.06$ & \cellcolor[HTML]{D3D3D3}$0.91 \pm 0.04$ & \cellcolor[HTML]{D3D3D3}$0.73 \pm 0.1$ & \cellcolor[HTML]{D3D3D3}$0.85 \pm 0.06$ \\
         & $\checkmark$ &  Hopfield  & $0.73 \pm 0.13$ & $0.92 \pm 0.29$ & $0.81 \pm 0.05$ & $0.9 \pm 0.03$ & $0.71 \pm 0.04$ & $0.84 \pm 0.03$\\
        \midrule
        
        \multirow{4}{*}{ResNet50}  & -& -& $4.51 \pm 1.11$ & $30.75 \pm 12.73$ & $0.08 \pm 0.05$ & $0.1 \pm 0.26$ & $0.09 \pm 0.04$ & $0.11 \pm 0.27$\\
         & $\checkmark$ & -& $0.96 \pm 0.28$ & $3.01 \pm 3.65$ & $0.63 \pm 0.25$ & $0.77 \pm 0.19$ & $0.54 \pm 0.23$ & $0.72 \pm 0.17$\\
         & $\checkmark$ & Transformer & $0.79 \pm 0.05$ & $0.98 \pm 0.12$ & \cellcolor[HTML]{D3D3D3}$0.83 \pm 0.06$ & \cellcolor[HTML]{D3D3D3}$0.91 \pm 0.03$ &\cellcolor[HTML]{D3D3D3} $0.64 \pm 0.13$ & \cellcolor[HTML]{D3D3D3}$0.8 \pm 0.08$\\
         & $\checkmark$ &  Hopfield  &\cellcolor[HTML]{D3D3D3} $0.77 \pm 0.09$ & \cellcolor[HTML]{D3D3D3}$0.96 \pm 0.16$ & $0.82 \pm 0.06$ & \cellcolor[HTML]{D3D3D3}$0.91 \pm 0.03$ & $0.62 \pm 0.12$ & $0.78 \pm 0.08$\\
        \midrule
        
        \multirow{4}{*}{EfficientNet-B1}  & -& -& $1.41 \pm 0.27$ & $3.52 \pm 1.1$ & $0.29 \pm 0.11$ & $0.52 \pm 0.11$ & $0.26 \pm 0.08$ & $0.51 \pm 0.08$\\
         & $\checkmark$ & -& $0.93 \pm 0.09$ & $1.39 \pm 0.28$ & $0.69 \pm 0.11$ & $0.83 \pm 0.07$ & $0.63 \pm 0.14$ & $0.79 \pm 0.08$\\
         & $\checkmark$ & Transformer&\cellcolor[HTML]{D3D3D3} $0.75 \pm 0.05$ & \cellcolor[HTML]{D3D3D3}$0.93 \pm 0.07$ & $0.79 \pm 0.06$ & \cellcolor[HTML]{D3D3D3}$0.89 \pm 0.04$ & \cellcolor[HTML]{D3D3D3}$0.71 \pm 0.09$ & \cellcolor[HTML]{D3D3D3}$0.84 \pm 0.06$\\
         & $\checkmark$ &  Hopfield & $0.78 \pm 0.11$ & $0.98 \pm 0.24$ &\cellcolor[HTML]{D3D3D3} $0.8 \pm 0.05$ &\cellcolor[HTML]{D3D3D3} $0.89 \pm 0.03$ & $0.69 \pm 0.09$ & $0.82 \pm 0.05$\\
        \midrule
        
        \multirow{4}{*}{EfficientNet-B2}  & -& -& $1.46 \pm 0.13$ & $3.37 \pm 0.65$ & $0.27 \pm 0.11$ & $0.5 \pm 0.11$ & $0.23 \pm 0.1$ & $0.46 \pm 0.1$\\
         & $\checkmark$ & -& $1.05 \pm 0.13$ & $1.72 \pm 0.38$ & $0.65 \pm 0.07$ & $0.81 \pm 0.04$ & $0.51 \pm 0.14$ & $0.71 \pm 0.1$\\
         & $\checkmark$ & Transformer& $0.8 \pm 0.12$ & $1.03 \pm 0.17$ & $0.8 \pm 0.05$ & $0.89 \pm 0.03$ & \cellcolor[HTML]{D3D3D3}$0.64 \pm 0.12$ & \cellcolor[HTML]{D3D3D3}$0.8 \pm 0.07$\\
         & $\checkmark$ &  Hopfield &\cellcolor[HTML]{D3D3D3} $0.78 \pm 0.09$ & \cellcolor[HTML]{D3D3D3}$0.99 \pm 0.18$ & \cellcolor[HTML]{D3D3D3}$0.81 \pm 0.06$ & \cellcolor[HTML]{D3D3D3} $0.9 \pm 0.03$ &\cellcolor[HTML]{D3D3D3} $0.64 \pm 0.1$ & \cellcolor[HTML]{D3D3D3} $0.8 \pm 0.06$\\
        
        \midrule
        \multirow{4}{*}{EfficientNet-B3}  & -& -& $1.59 \pm 0.27$ & $4.28 \pm 1.26$ & $0.11 \pm 0.09$ & $0.31 \pm 0.12$ & $0.18 \pm 0.14$ & $0.37 \pm 0.2$\\
         & $\checkmark$ & -& $1.15 \pm 0.06$ & $2.21 \pm 0.45$ & $0.58 \pm 0.12$ & $0.76 \pm 0.08$ & $0.53 \pm 0.11$ & $0.73 \pm 0.09$\\
         & $\checkmark$ & Transformer& \cellcolor[HTML]{D3D3D3}$0.68 \pm 0.02$ & \cellcolor[HTML]{D3D3D3} $0.76 \pm 0.07$ & \cellcolor[HTML]{D3D3D3} $0.85 \pm 0.05$ & \cellcolor[HTML]{D3D3D3} $0.92 \pm 0.02$ & \cellcolor[HTML]{D3D3D3} $0.72 \pm 0.09$ & \cellcolor[HTML]{D3D3D3} $0.85 \pm 0.06$\\
         & $\checkmark$ &  Hopfield & $0.74 \pm 0.1$ & $0.93 \pm 0.14$ & $0.82 \pm 0.05$ & $0.9 \pm 0.03$ & $0.7 \pm 0.11$ & $0.84 \pm 0.06$\\
        \midrule
        \multirow{4}{*}{EfficientNet-B4}  &- &- & $1.96 \pm 0.36$ & $6.29 \pm 1.87$ & $0.13 \pm 0.07$ & $0.34 \pm 0.1$ & $0.15 \pm 0.06$ & $0.38 \pm 0.09$\\
         & $\checkmark$ &- & $1.08 \pm 0.21$ & $1.77 \pm 0.62$ & $0.62 \pm 0.11$ & $0.79 \pm 0.07$ & $0.5 \pm 0.1$ & $0.7 \pm 0.07$\\
         & $\checkmark$ &Transformer & \cellcolor[HTML]{D3D3D3}$0.81 \pm 0.14$ & \cellcolor[HTML]{D3D3D3} $1.04 \pm 0.33$ & \cellcolor[HTML]{D3D3D3} $0.81 \pm 0.08$ & \cellcolor[HTML]{D3D3D3} $0.9 \pm 0.04$ & \cellcolor[HTML]{D3D3D3} $0.63 \pm 0.09$ & $0.79 \pm 0.05$\\
         & $\checkmark$ &  Hopfield  & \cellcolor[HTML]{D3D3D3}$0.81 \pm 0.09$ & $1.18 \pm 0.2$ & $0.75 \pm 0.06$ & $0.86 \pm 0.04$ & \cellcolor[HTML]{D3D3D3} $0.63 \pm 0.07$ & \cellcolor[HTML]{D3D3D3} $0.8 \pm 0.04$\\
        \bottomrule
    \end{tabular}
    \caption{Experimental results of COVID-19 severity prediction. Each block report the performance of individual backbone CNNs under four experimental setting. The global evaluation metrics are MAE and MSE, while evaluation metrics for individual attributes are $R^2$ and Pearson correlation coefficient $\rho$. The best results withing each box are shaded in gray. The same training setting is used across all experiments, which is reported in Section~\ref{sec:severity-prediction}}
    \label{tab:main-results}
\end{table*}
\subsubsection{COVID-19 severity prediction}\label{sec:severity-prediction}
The COVID-19 dataset for severity prediction is from \cite{cohen2020predicting}, which is completely separated from the proxy dataset. The database includes $94$ posteroanterior (PA) chest X-ray images. All patients in the cohort have been reported positive to COVID-19 from December 2019 to March 2020. The labels of the database are based on radiological scoring, which involved three blinded experts. Two chest radiologist (with 20 years of experience) and a radiology resident score the COVID-19 severity based on \cite{wong2020frequency}, which includes extent of lung involvement (\textbf{geographic extend}) and degree of opacity (\textbf{opacity}).

The COVID-19 severity prediction dataset can be considered an extremely small dataset, so we decided to perform 5-fold cross-validation to evaluate competitors' performance. First, we split the dataset into five independent folds, which guarantee no overlapped patient across folds. The evaluation metrics for severity prediction are: (1) mean squared error (MSE), (2) mean absolute error (MAE), (3) R-squared, and Pearson correlation between actual and predicted scores.

The loss function used in this phase is smoothed L1 loss with $\beta = 1$, which is given by
\begin{equation}\label{smoothL1}
    \mathcal{L} (\bm{y,\hat{y}}) = \begin{cases}
    0.5 (\bm{y} - \bm{\hat{y}})^2 /\beta \\ 
    |\bm{y} - \bm{\hat{y}}| - 0.5*\beta 
    \end{cases}
\end{equation}
We train the model on each fold for $400$ epochs with the SGD optimizer with an initial learning rate of $10^{-3}$, momentum $0.9$, and weight decay $3\times 10^{-5}$. In order to prevent the over-fitting problem, we reduce the learning rate with a decay rate of $0.98$ for every $2$ epoch, and the dropout rate for the self-attention module is set at $0.1$. 

\subsection{Pre-training Results}

Figure~\ref{fig:pretraining} depicts the performances of five investigating backbone CNNs, in terms of multi-label classification. The mean area under curve (AUC) scores of Densenet121, EfficientNet-B1, B2, and B3 are nearly the same, while EfficientNet-B4 achieve a slightly higher AUC of $0.8018$ on $15$ classes. On the other hand, ResNet50 achieves the least AUC score of $0.7495$, although it has the largest model complexity. 

Table~\ref{tab:pretraining} reports the AUC score for each class from five CNNs models. As we can see, the AUC score of the COVID-19 class is very close to perfect prediction. However, the phenomenon is potentially attributed to the cross-domain design of proxy dataset, where COVID-19 images are from entirely different institutions \cite{cohen2020limits}. Thus, we do not attempt to compare the results of the pre-training phase to other works or emphasize the COVID-19 detection ability. Instead, the main objective of the pre-training phase is assisting backbone models to learn valuable representations for the downstream task, which is severity prediction. We will investigate the effectiveness of such knowledge transfer and expansion here in Section~\ref{experiment}.

\subsection{Severity Prediction}

\subsubsection{Effects of Data-centric pre-training}

We report the main results of the COVID-19 severity prediction task in Table~\ref{tab:main-results}, which contains six blocks corresponding to the choices of backbone CNN.

In the first line of each block, we report the performance of stand-alone CNNs under transfer learning setup, in which model's weights are inherited from ImageNet. The outcomes are consistent from all backbone CNNs, showing that transfer learning is \textit{ineffective} in the case of extremely small and irrelevant target datasets. The performance on the test set of higher complexity such as ResNet50 and EfficientNet-B4 is lower than small neural networks even though over-fitted on the training set. Moreover, from the diagnosis of learning curves (not shown here), these models stop gaining test accuracy after half of the training process, even being applied over-fitting prevention such as adaptive learning rate or dropout.

In the second line, we train stand-alone backbone CNNs with weights from our data-centric pre-training task. The consistent pattern appears across all models, yielding improvement noticeably in comparison to transfer learning from ImageNet. First, the test accuracy is improved significantly, which can be observed through test MAE and MSE. Moreover, the $R^2$ and Pearson correlation between actual and predicted values increases with a considerably large gap, indicating a more precise prediction. The most accuracy gain can be observed from DenseNet121, while deeper CNNs such as ResNet50 gain a minor improvement. However, the agreement between actual and predicted is not remarkable, which achieves only $R^2$ of $0.74$ and $0.55$ to predict geographic extend and opacity from DenseNet121.

Figure~\ref{fig:training} illustrates the cross-validation loss computed by Equation~\ref{smoothL1}. Transfer learning fails to achieve good performance in comparison to data-centric pre-training.

\subsubsection{Effects of Hybrid Convolution-Attention Architecture}

The third and fourth line of Table~\ref{tab:main-results} shows the performance of proposed hybrid convolution-attention architectures with different self-attention modules. We denote HCT for Transformer and HCH for Hopfield network. The initial weights for these experiments are adopted from the data-centric pre-training phase. Overall, the MSE of hybrid neural architectures drops approximately $0.2$ points across all backbone models, while MAS drops $1.5$ points on average. We can see the significant improvement when observing the $R^2$ from each hybrid model. For example, the $R^2$ in geographic extend from DenseNet121 increase from $0.74$ to $0.82$, while that in opacity prediction enhances from $0.55$ to $0.73$. Moreover, Figure~\ref{fig:pretraining} depicts that the CV-loss of hybrid architectures is lower than stand-alone CNNs in general, while the difference between two self-attention modules is not noticeable. We report the alignment of DenseNet121, HCT-DenseNet121 and HCH-DenseNet121 in Figure~\ref{fig:densenet121}. In general, hybrid architectures achieve better performance across all folds. Moreover, the alignment from geographic extent is slightly better than the predictions of opacity.

\begin{figure*}[!t]
    \centering
    \includegraphics[width = 0.3\textwidth]{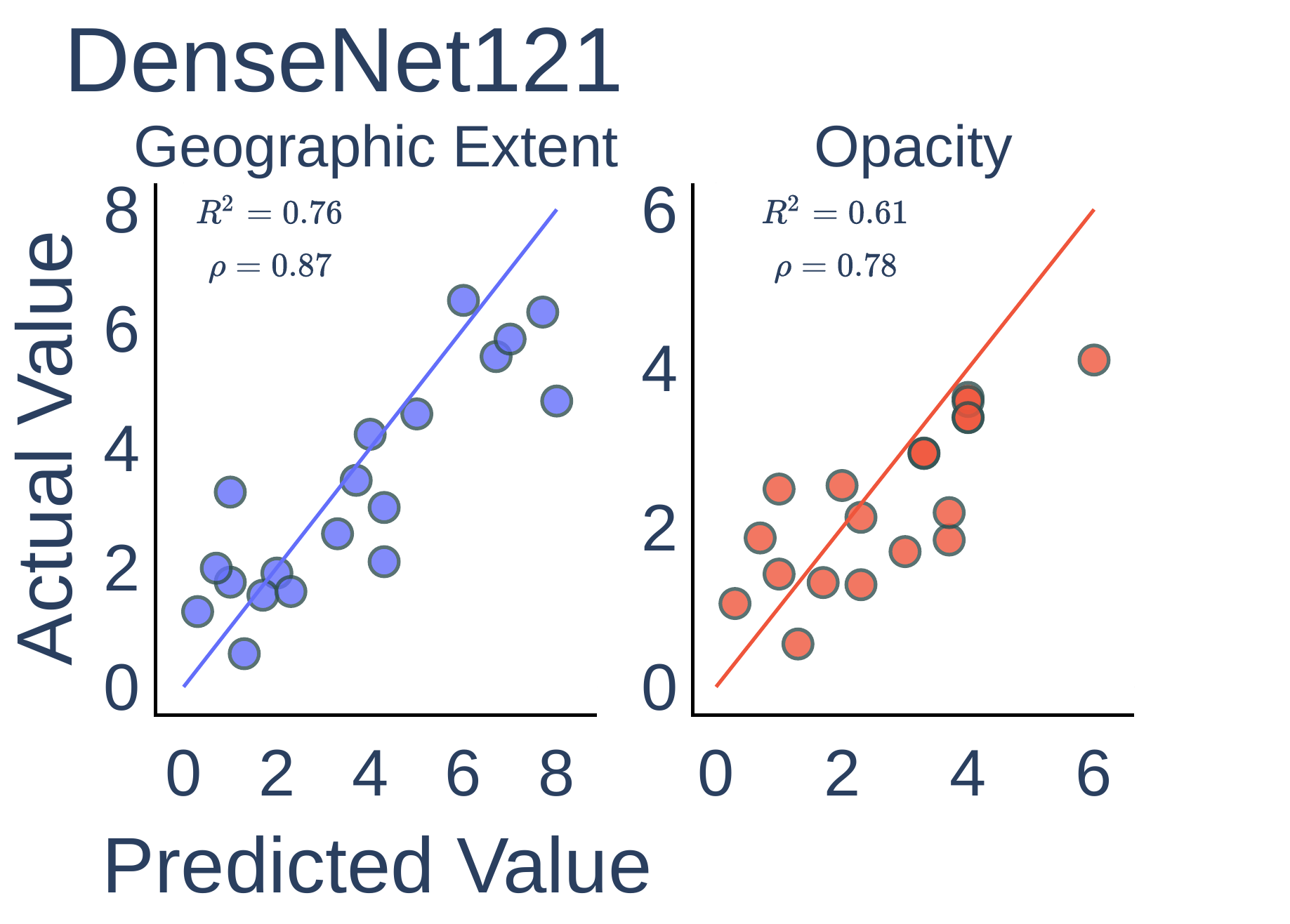}
    \includegraphics[width = 0.3\textwidth]{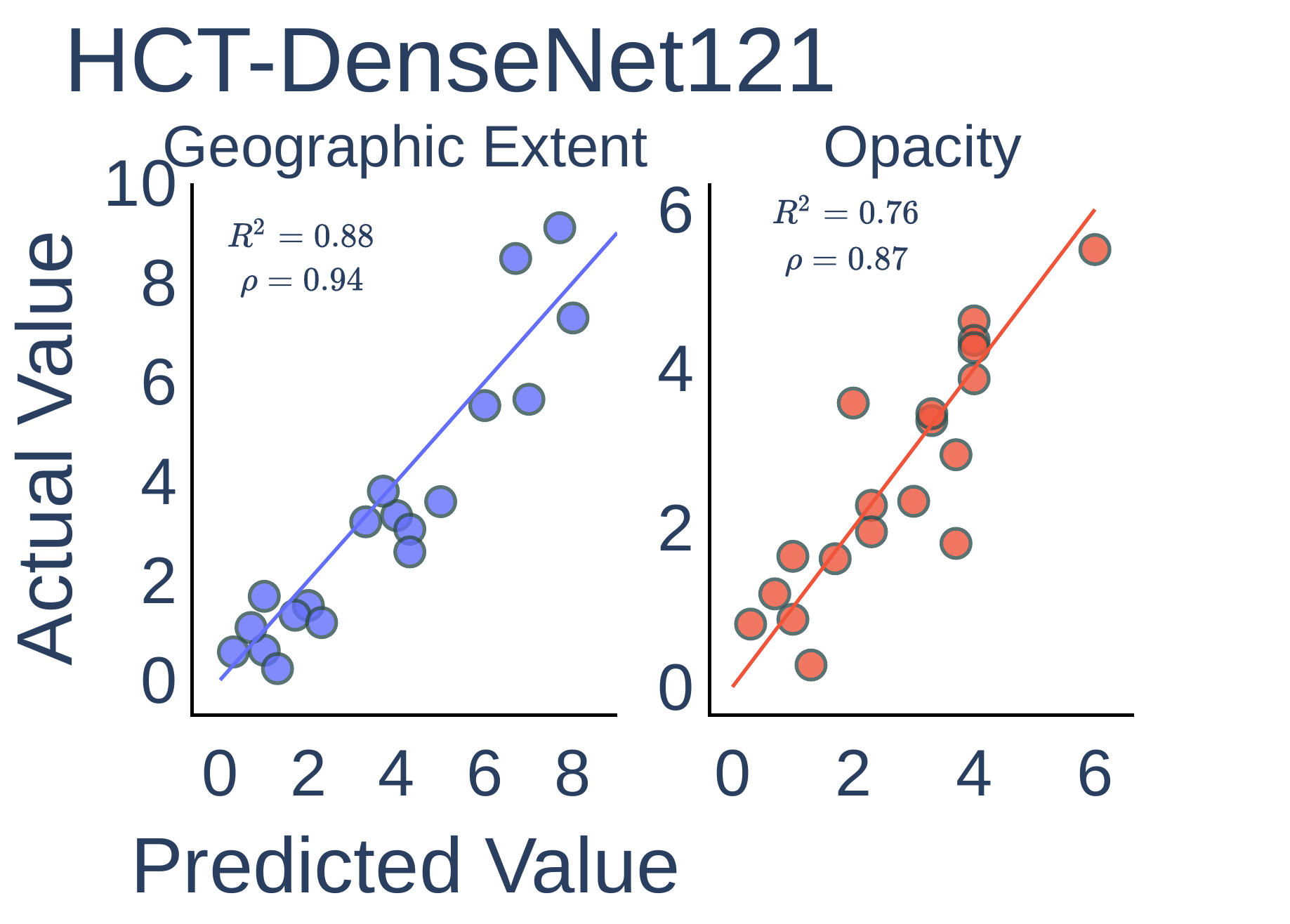}
    \includegraphics[width = 0.3\textwidth]{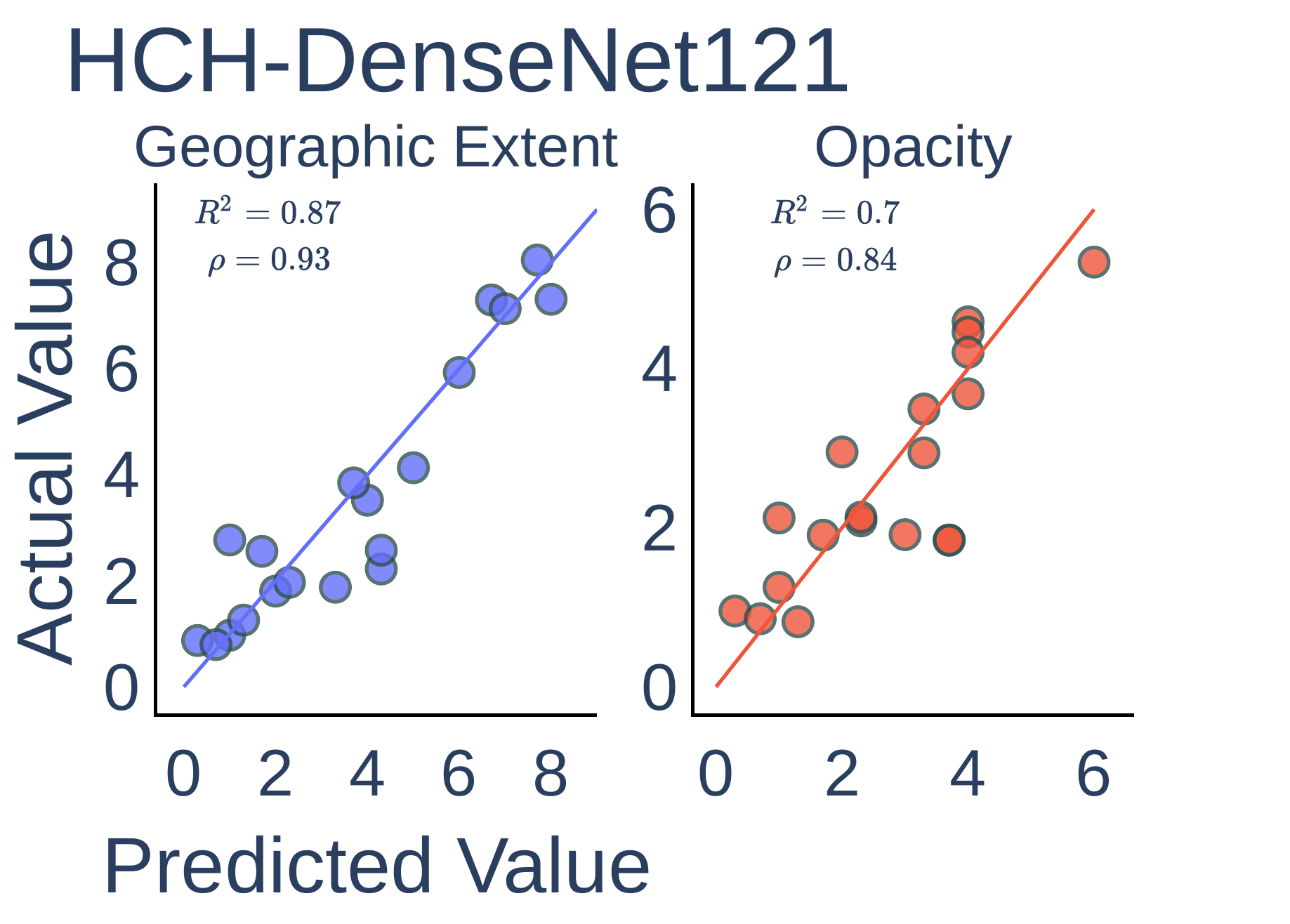}
    \includegraphics[width = 0.3\textwidth]{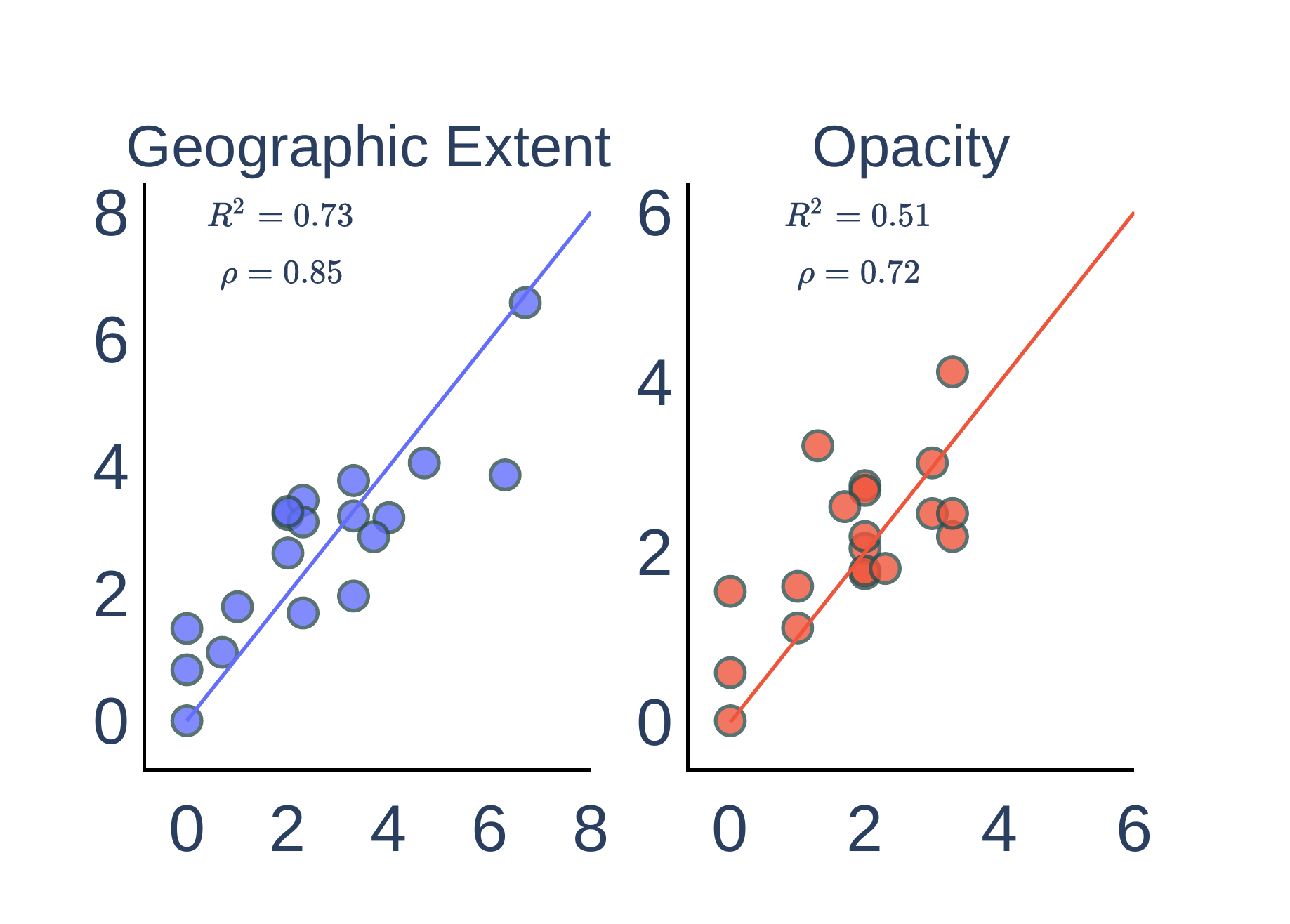}
    \includegraphics[width = 0.3\textwidth]{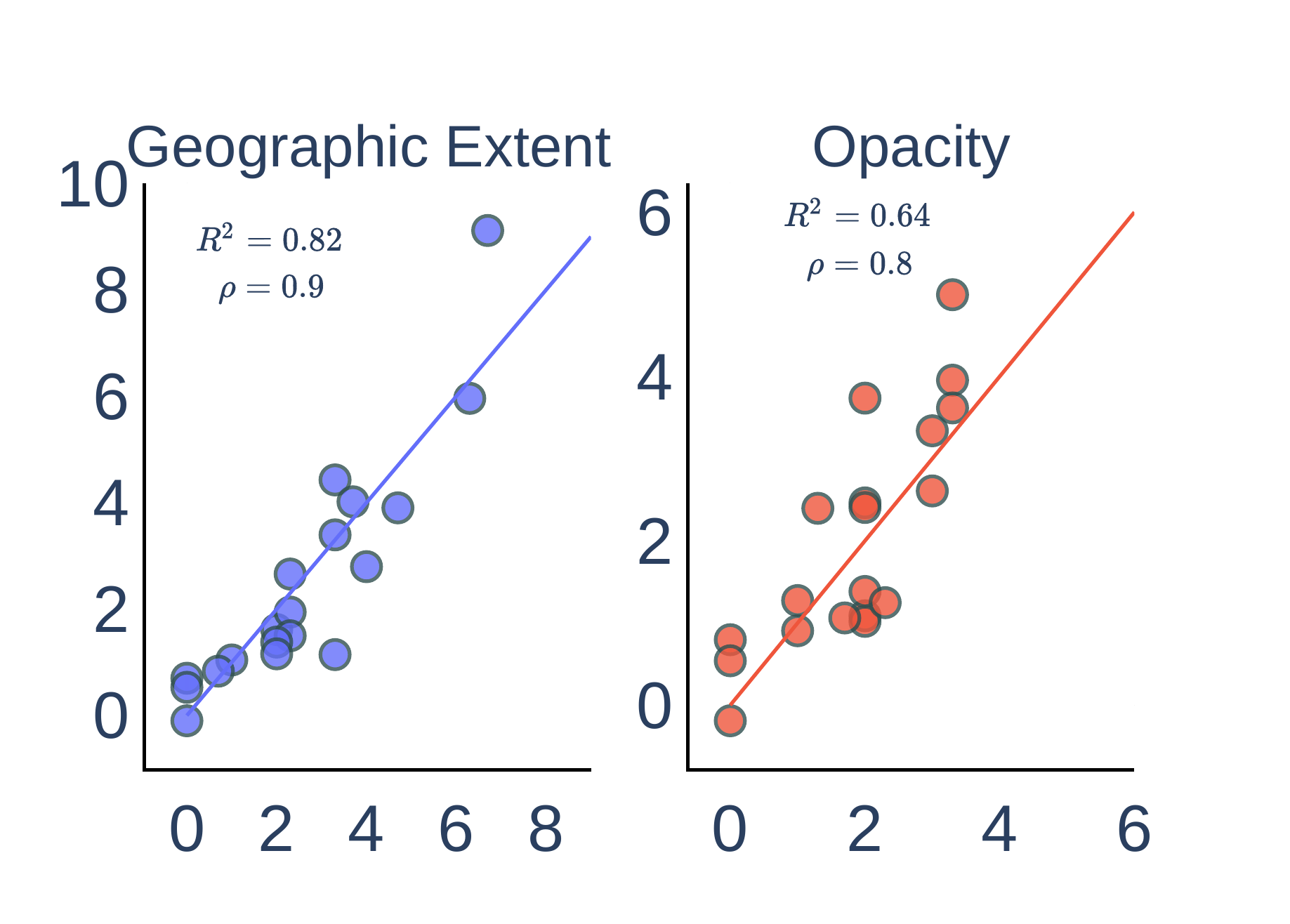}
    \includegraphics[width = 0.3\textwidth]{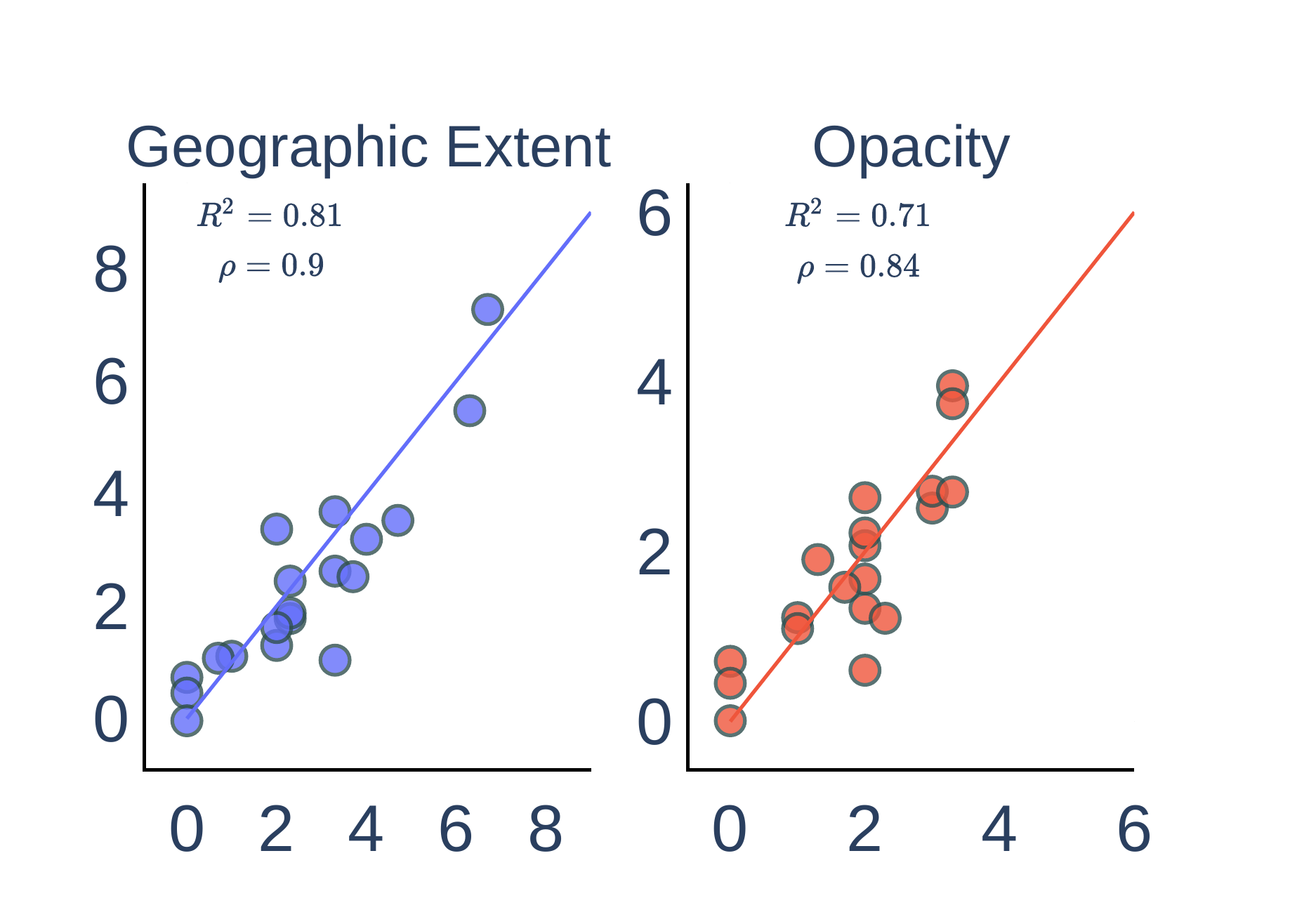}
    \includegraphics[width = 0.3\textwidth]{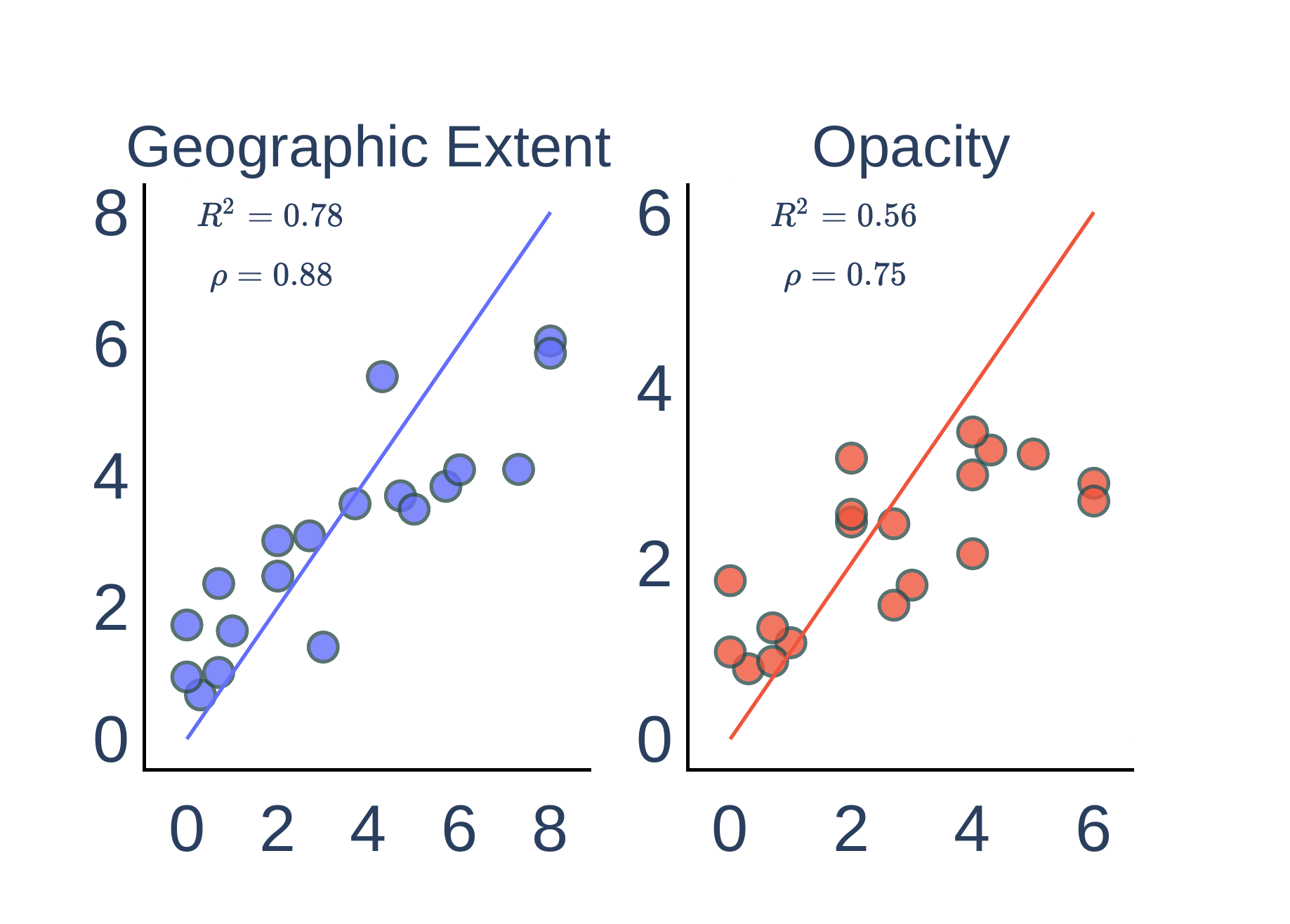}
    \includegraphics[width = 0.3\textwidth]{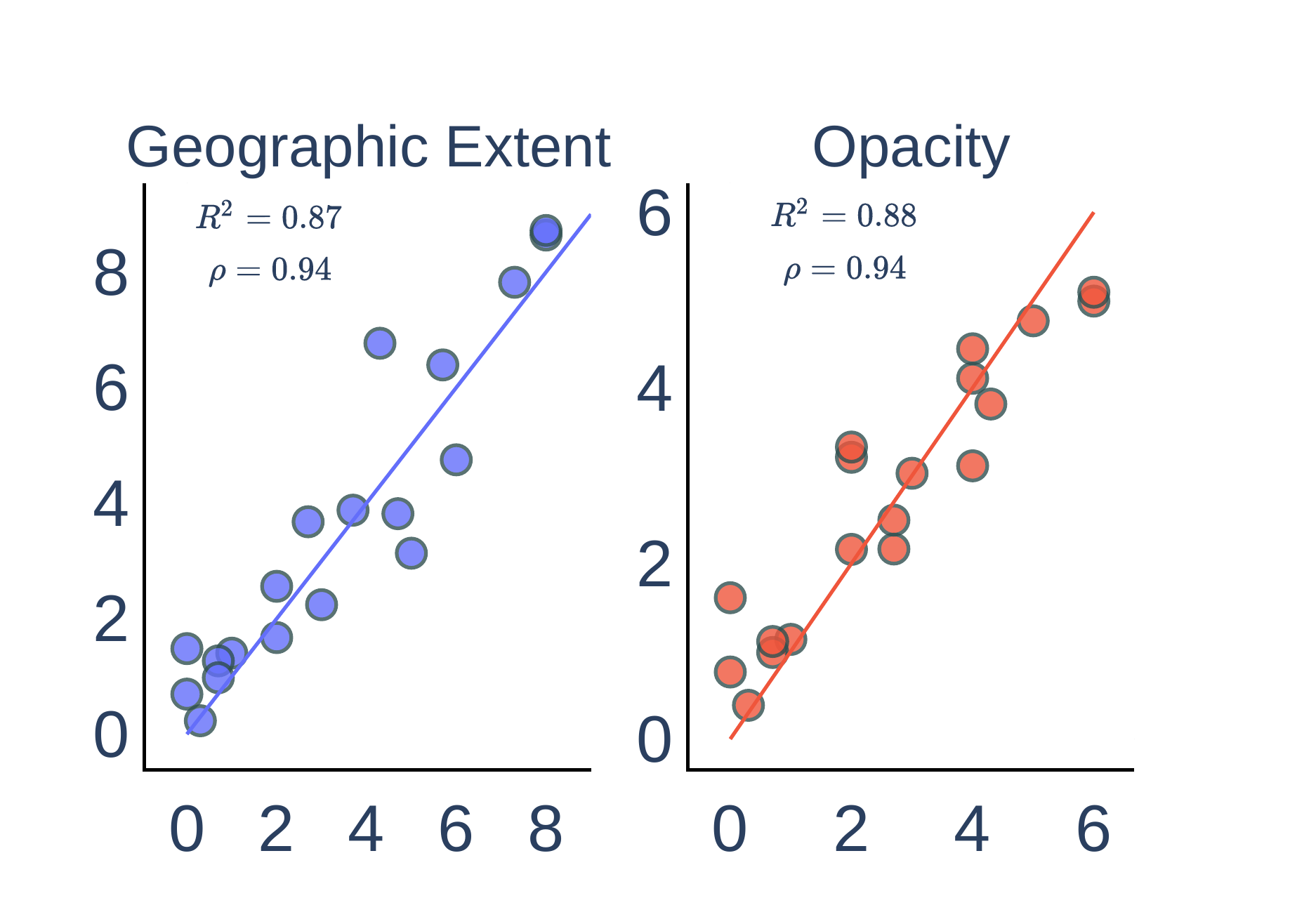}
    \includegraphics[width = 0.3\textwidth]{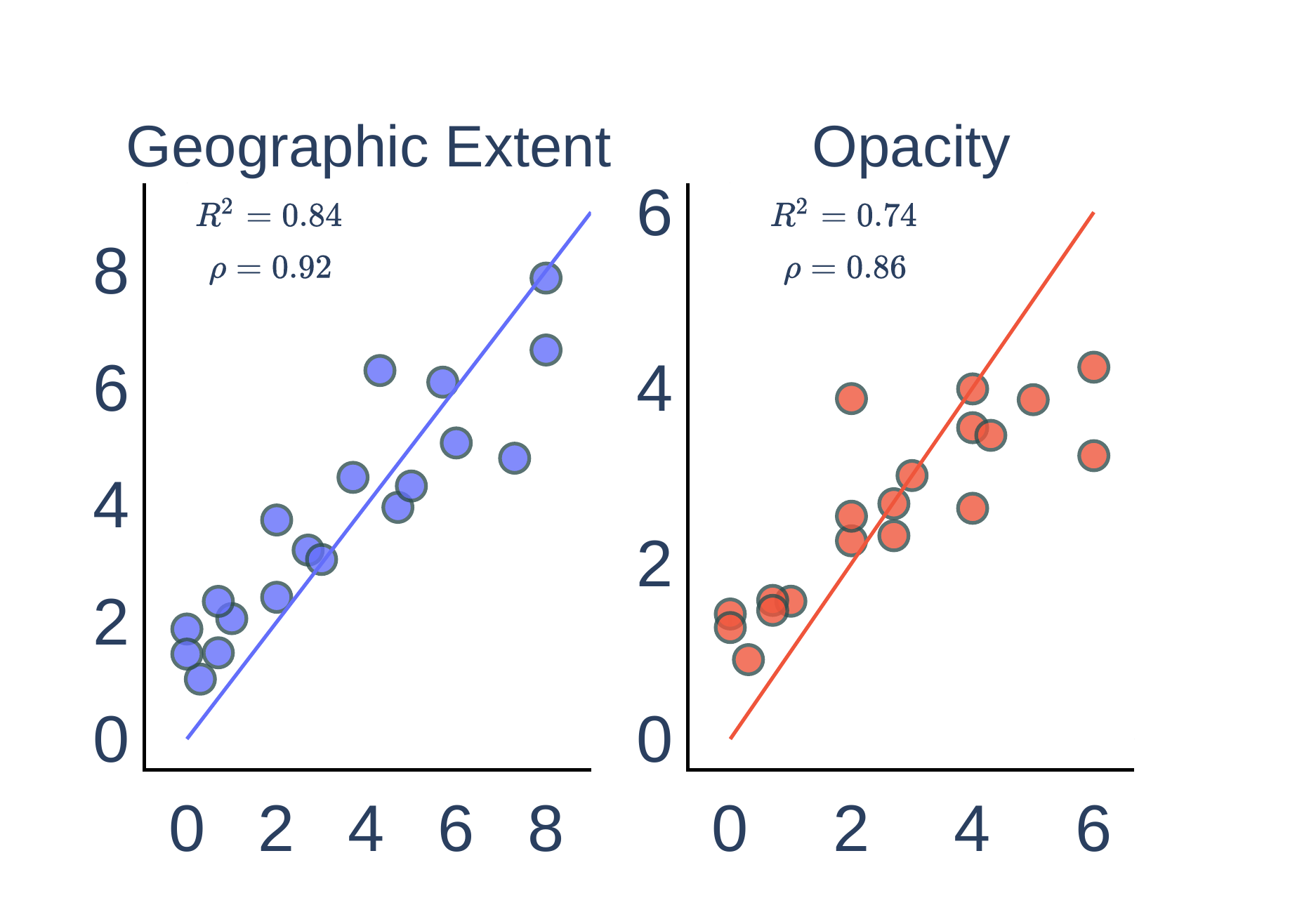} 
    \includegraphics[width = 0.3\textwidth]{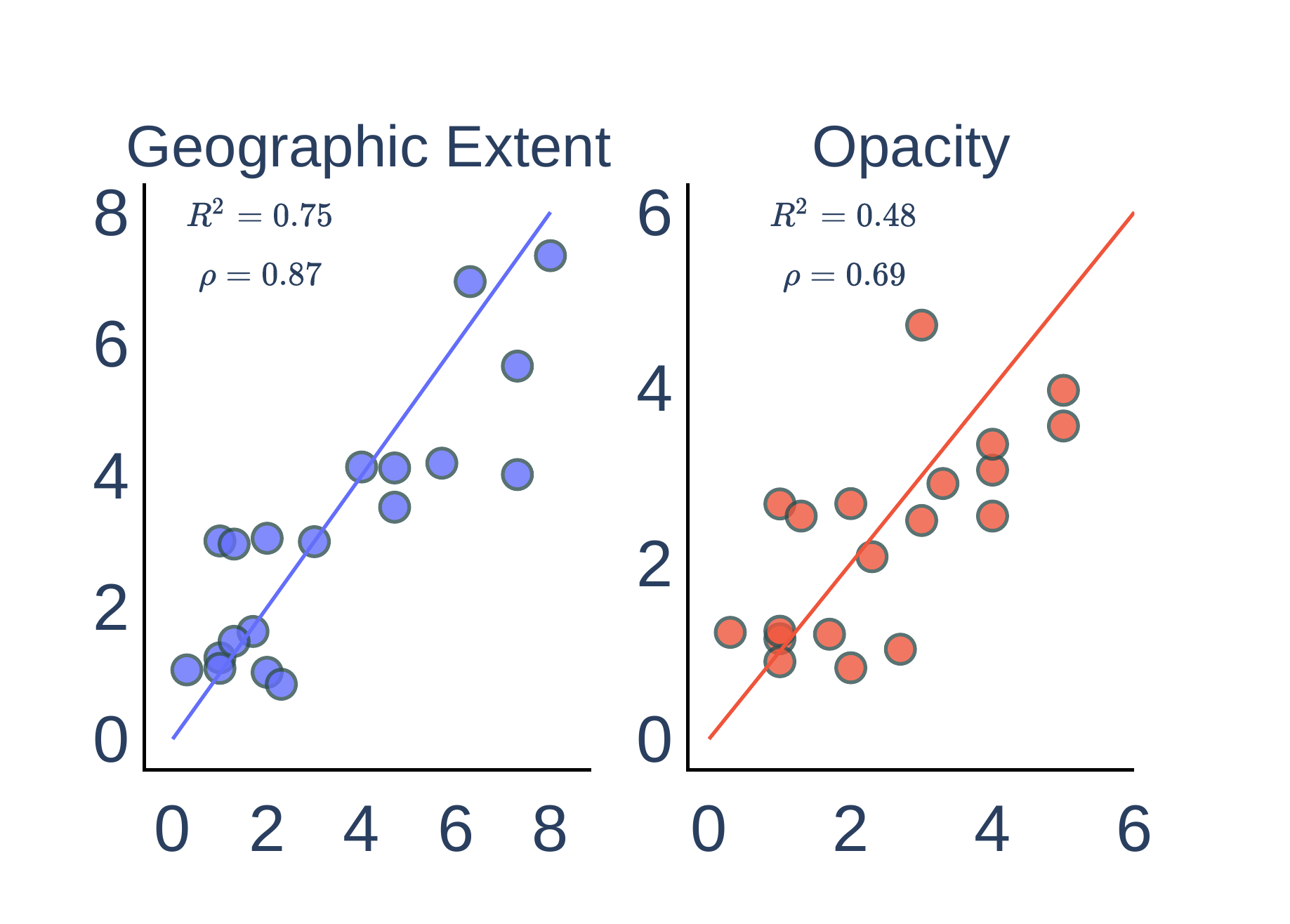}
    \includegraphics[width = 0.3\textwidth]{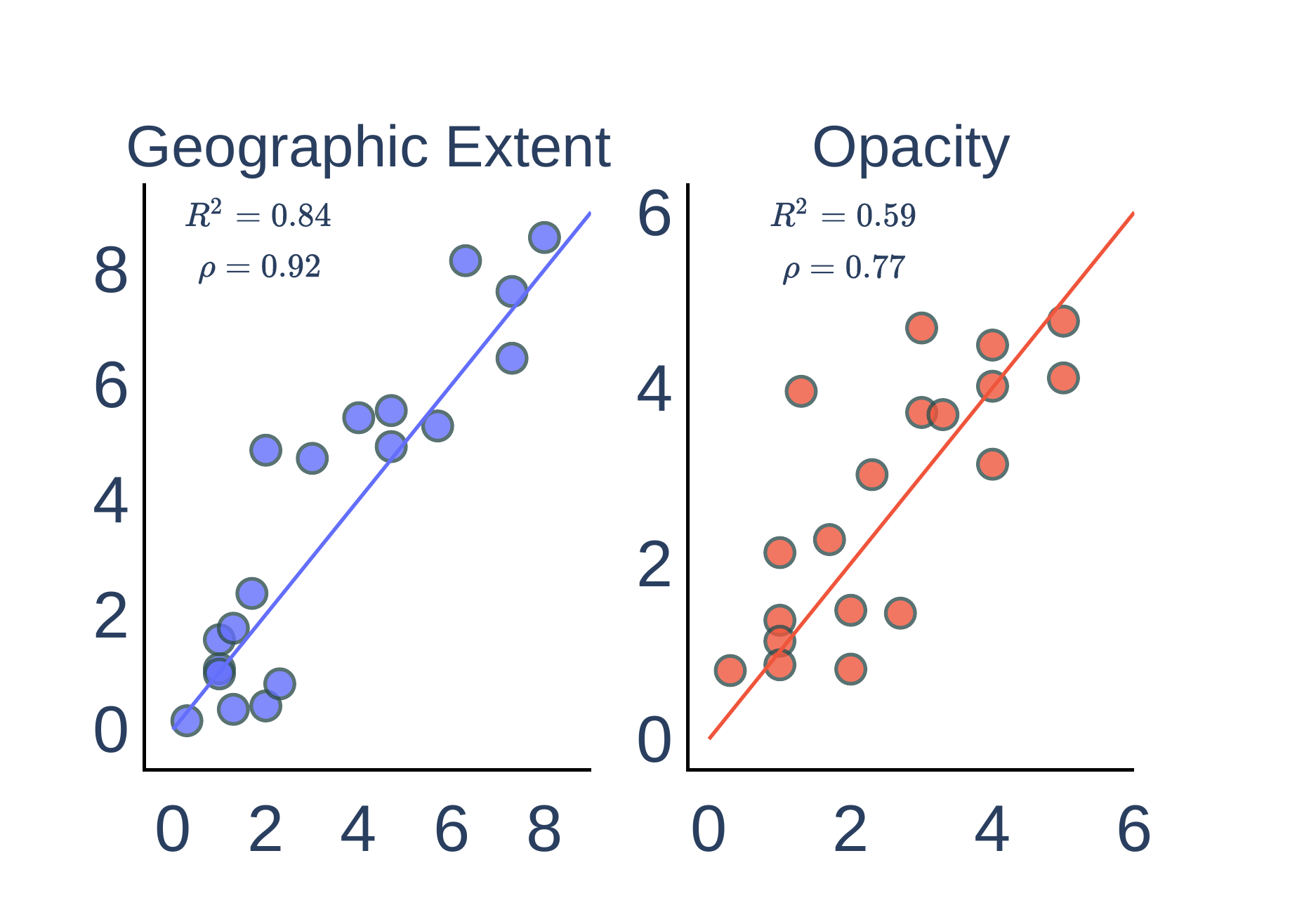}
    \includegraphics[width = 0.3\textwidth]{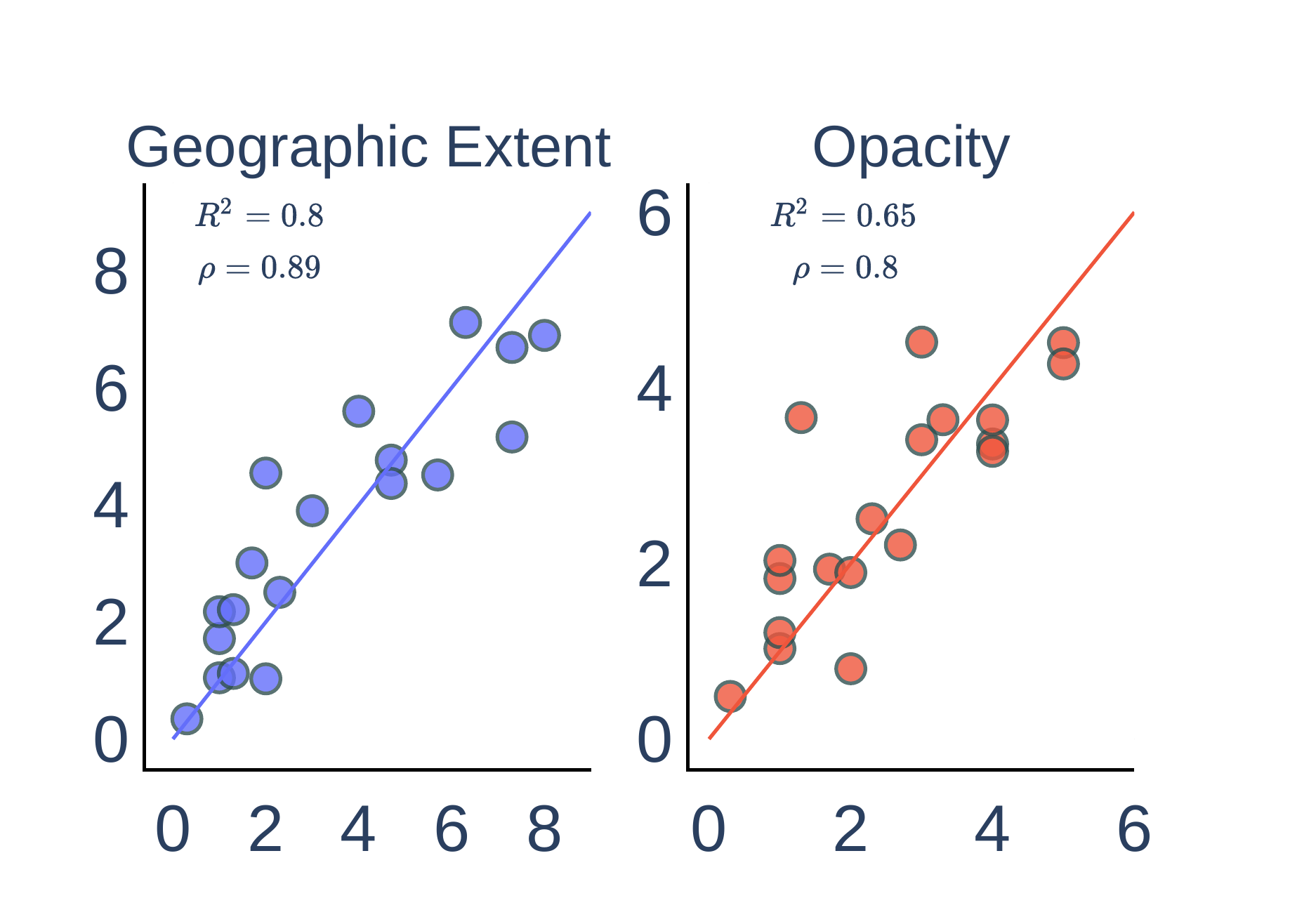}
    \includegraphics[width = 0.3\textwidth]{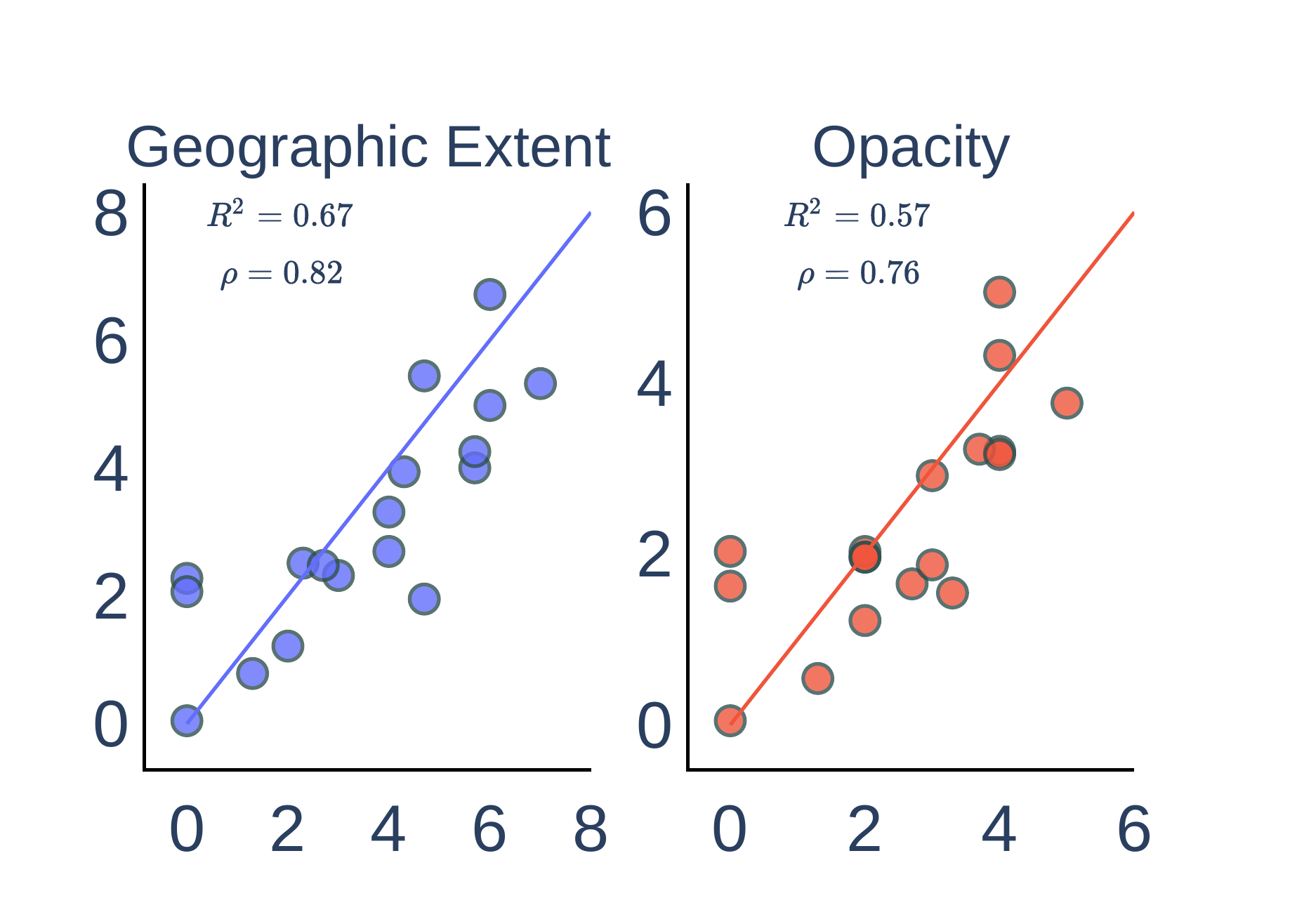}
    \includegraphics[width = 0.3\textwidth]{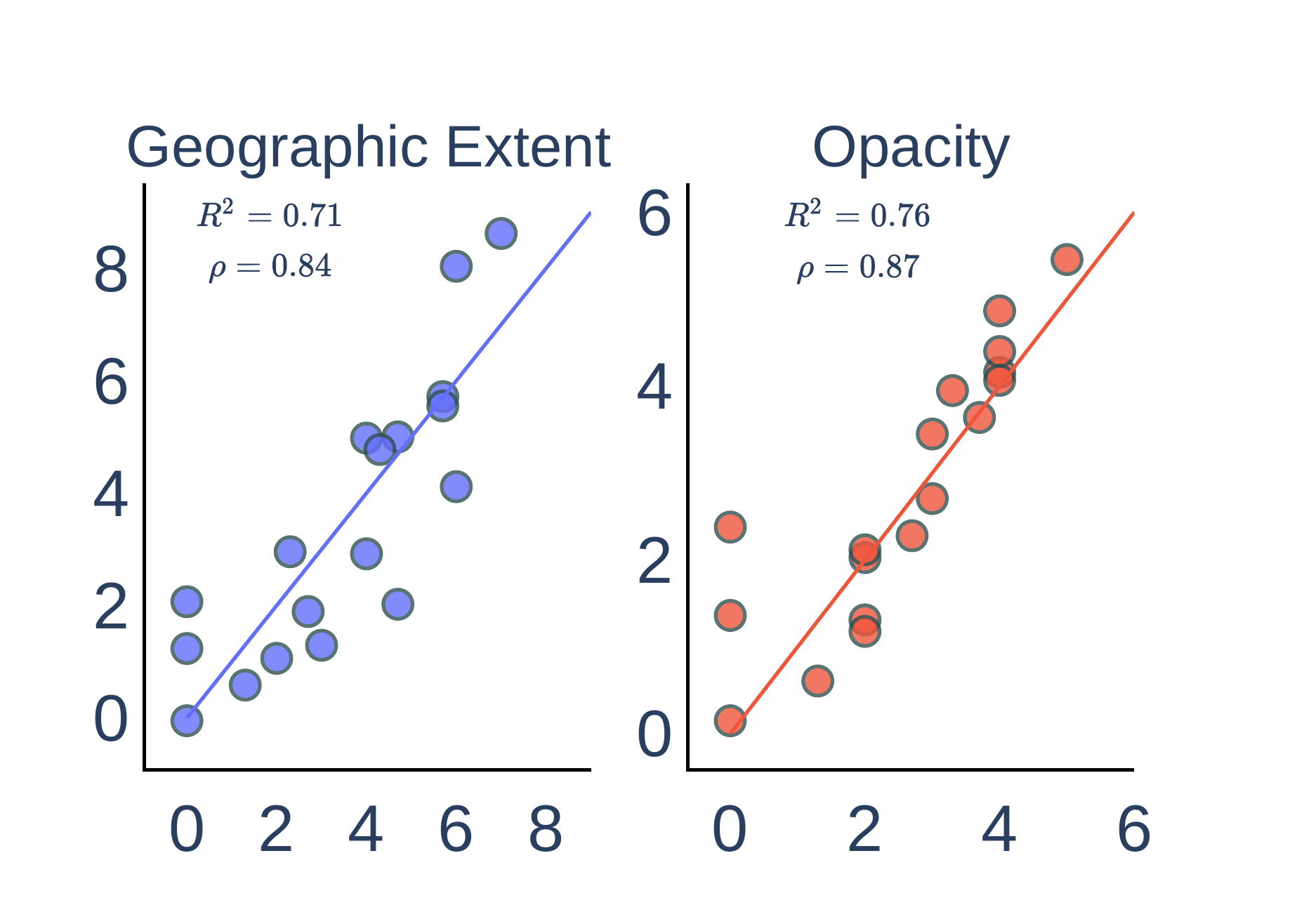}
    \includegraphics[width = 0.3\textwidth]{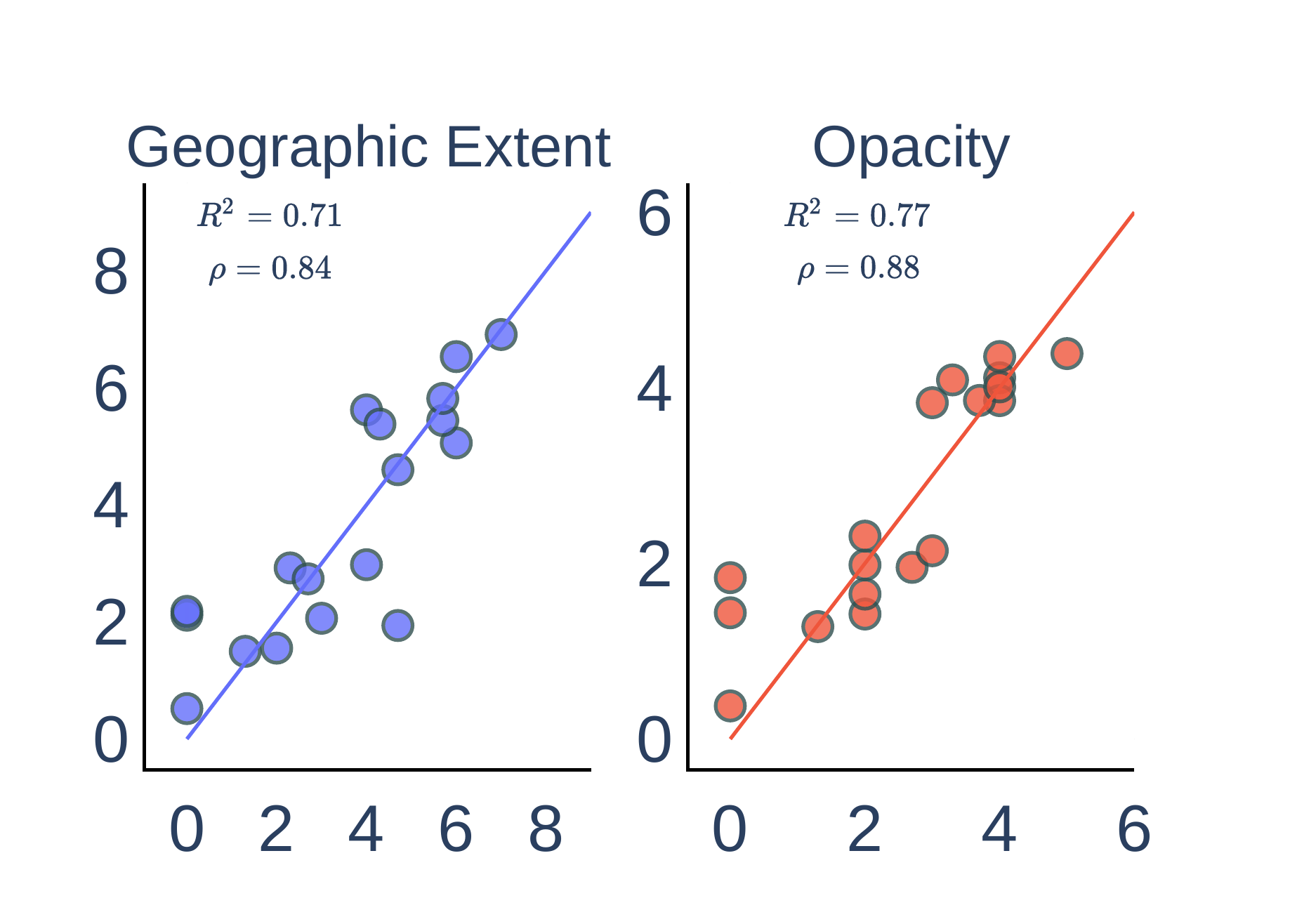}
    \caption{An exemplar alignment between actual and predicted values from the experiment on DenseNet121. Each column shows different investigating models, while each row shows the results on each fold. The solid straight line indicates perfect prediction, and $\rho$ is the Pearson correlation coefficient.}
    \label{fig:densenet121}
\end{figure*}

\section{Conclusion}\label{conclusion}

We have presented a novel framework for COVID-19 severity prediction. Our data-centric pre-training design enables high performance models when transferring knowledge to the downstream task. Moreover, we introduce new class of deep neural architecture, which capture the global contextual information from the input space through self-attention modules. Further improvement of our work considers different self-attention modules for the hybrid architecture. Additionally, extending the framework to more applications is also a potential research direction.

\section*{Acknowledgments}
Effort sponsored in part by United States Special Operations Command (USSOCOM), under Partnership Intermediary Agreement No. H92222-15-3-0001-01. The U.S. Government is authorized to reproduce and distribute reprints for Government purposes, notwithstanding any copyright notation thereon. \footnote{The views and conclusions contained herein are those of the authors and should not be interpreted as necessarily representing the official policies or endorsements, either expressed or implied, of the United States Special Operations Command.}

\bibliography{output}
\bibliographystyle{IEEEtran}

\end{document}